\documentclass[preprint,1p,11pt]{ISAS_IR}

\usepackage{ucs} 
\usepackage[utf8x]{inputenc}
\usepackage{pstool}
\usepackage{subfig}

\usepackage{ifthen}

\usepackage{ISASTextMacros}
\usepackage{ISASMatheMacros}

\newsavebox\myVerb

\newcommand*{\verbBox}{\usebox\myVerb}
\newcommand{\placeFig}[1]{}

\renewcommand{\psfragfig}{\includegraphics}

\begin{document}

\begin{frontmatter}

\title{Extended Object Tracking with \\Random Hypersurface Models}
\author{%
\textbf{Marcus Baum} and \textbf{Uwe D.~Hanebeck}\\
Intelligent Sensor-Actuator-Systems Laboratory (ISAS)\\
Institute for Anthropomatics\\
Karlsruhe Institute of Technology (KIT), Germany\\
{\tt marcus.baum@kit.edu, \tt uwe.hanebeck@ieee.org}
}
\begin{abstract}
The \RHM\ (\rhm) is introduced that allows for  estimating a shape approximation of an extended object in addition to its  kinematic state. An \emph{RHM} represents the spatial extent by means of randomly scaled versions of the shape boundary.  In doing so,  the shape parameters and the measurements are related via a measurement equation that serves as the basis for a  Gaussian state estimator. Specific estimators  are derived for elliptic and star-convex shapes.
\end{abstract}\tnotetext[t1]{Draft accepted for publication in IEEE Transactions on Aerospace and Electronic Systems.}
\end{frontmatter}

\chapter{Introduction}\label{sec:intro}
\begin{figure*}
\center
 \includegraphics[page=1] {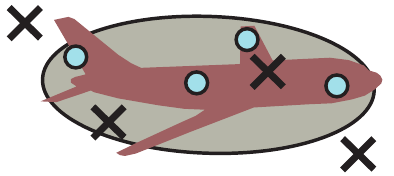}
 \hspace{0.2cm}
 \includegraphics[page=5,scale=0.8]  {Figures/Scenarios.pdf}
\caption{Shape approximation  of an extended object  with an ellipse\label{fig:scenarios_approx}. \label{fig:scenarios_approx1}}
\end{figure*}

In target tracking applications \cite{Bar-Shalom2002} where the resolution of  the  sensor device is higher than the spatial extent of a target object, the usual point object assumption is not justified as 
several different points, i.e., measurement sources, on the target object may  be  resolved during a  single scan (see \Fig{fig:scenarios_approx1}). The resolved measurement sources typically vary from scan to scan and their locations  depend on the shape of the object but also on further  properties such as  the surface or the target-to-sensor geometry.
\placeFig{1} 
In this article, the  basic idea is to approximate an extended object with a geometric shape such as an ellipse \cite{Gilholm2005,Koch2008,Fusion10_BaumNoack} as depicted in \Fig{fig:scenarios_approx1}.
The tracking problem then consists of estimating the  shape parameters in addition to the kinematic parameters.
The locations of the measurement sources are not explicitly estimated.
Reasonably, the shape of the target should be described as detailed as possible. However, when the measurement noise is rather high  and only a few measurements are available, it may only be possible to infer a  coarse shape approximation such as a circle.  

The unknown locations of the measurement sources  are usually  modeled with  a probability distribution  whose mass is concentrated on the extended object (a so-called spatial distribution) \cite{Gilholm2005,G2005}.
In general, no closed-form solutions  for the likelihood function resulting from a spatial distribution model exist, so that Monte Carlo methods are frequently used for approximating the Bayesian filter solution \cite{Gilholm2005,G2005,Boers2006,Angelova2008,Petrov2011,Petrov2012a,Petrov2012}. 
In case an elliptic extent is represented with a covariance matrix  \cite{Koch2008,Feldmann2010,Degerman2011,Lan2012,Lan2012a,Wieneke2012}, closed-form expressions can be derived with the help of random matrix theory.
 Spatial distributions have been embedded into the \emph{Probability Hypothesis Density (PHD)} filter for tracking multiple extended objects  \cite{Mahler2009,Granstrom2010,Orguner2011,Lundquist2011a,Granstrom2012}.   
In \cite{Fusion09_Baum,TAES_Baum} it is proposed to drop all statistical assumptions on the measurement sources, which leads to combined set-theoretic and stochastic estimator.

\section{Contributions}
The main contribution  is a novel systematic approach for  modeling the  unknown location of a measurement source on a spatially  extended object  called \RHM\ (\rhm).
The basic idea is to  assume the measurement source  to lie on a scaled version of the shape boundary, where the scaling factor is modeled as a random variable.
In this manner, it is possible to  form a suitable measurement equation that serves as the basis for constructing  a Gaussian state estimator.  
In order to illustrate the novel approach, specific \rhms\ and corresponding Gaussian estimators are developed for ellipses and free-form star-convex shapes. 
To the best of our knowledge,   this is the first extended object tracking method for  \emph{explicitly} estimating a free-form star-convex shape approximation. Actually, with this method, it is possible to track a target  whose shape is a priori  unknown and estimated from scratch over time.

\begin{Remark} This article is based on   \cite{SDF10_Baum,Fusion11_Baum-RHM,ISSPIT09_Baum,Fusion10_BaumNoack,CDC11_Baum,Fusion10_BaumKlumpp,IROS11_Baum,SYSAES_Baum,Diss13_Baum}.
\end{Remark}
\section{Overview}

The remainder of this article is structured as follows:
In the following section, the general probabilistic framework for extended object tracking is presented.
The new target extent model called \RHM\ (\rhm)  is subsequently introduced in    \Sec{sec:rhm}.
Based on these models, a formal Bayes filter for extended object tracking is described in \Sec{sec:bayes_est}.
Then,  particular implementations of \rhms\ for  ellipses (\Sec{sec:rhm_ellipse}) and star-convex shapes (\Sec{sec:rhm_starconvex}) are developed.
Both shape representations are evaluated by means of typical extended object tracking scenarios in \Sec{ssec:eval_starconvex}. This article is concluded in \Sec{sec:conclusions}.

\chapter{Modeling Extended Targets}
The  state vector of the extended object at discrete time  $k$  is represented with a random vector\footnote{In this article vectors are underlined, e.g., $\vec{x}$ is a vector,  and random variables are written in bold, e.g., $\rvec{x}$  denotes a random vector.} $\xAll{k}=\tvect{ \rvec{m}^T_k, (\rvec{x}^*_k)^T, \xShape{k}[T] }$ that consists of the target location $\rvec{m}_k$, a shape parameter vector $\xShape{k}$,  and an optional  vector $\rvec{x}^*_k$ for the kinematics, e.g., velocity.
The shape parameter $\xShape{k}$  specifies a two-dimensional set that  is denoted with $\shapeSet{\xShape{k}} \subset \IR^2$.
For example, a circular shape can be specified by its radius $\rv{r}_k$, i.e.,  $\rv{p}^\text{ci}_k = \rv{r}_k$  and  the corresponding  shape is  $\shapeSet{ \rv{p}^\text{ci}_k}=\{ \vec{z}  \; | \;    \vec{z} \in \IR^2 \text{ and } ||\vec{z}||_2 \leq  \rv{r}_k \}$.\footnote{The $\text{ci}$ in  $\rv{p}^{\text{ci}}_k$ indicates that it parameterizes a circle.}
Note that we focus on two-dimensional shapes. Nevertheless the concepts are also applicable to higher dimensional shapes in the same manner, see for example \cite{Fusion12_Faion-CylinderTracking}.

\section{Measurement Model}\label{sec:measmodel}
 \begin{figure}
\begin{center}
 \includegraphics {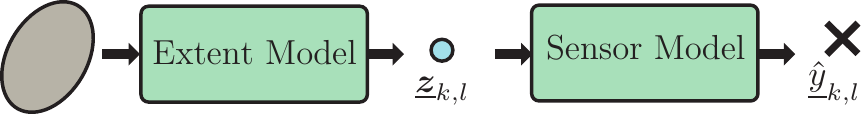}
\end{center}
\caption{Measurement model for a single measurement.  \label{fig:meas_model_sequ}}
\end{figure}
At each time step $k$, a set of  $n_k$ two-dimensional point measurements  $\{ \meas{k}[l]\}_{l=1}^{n_k}  $ of the extended object is available. 
We assume the measurements to be mutually independent for given target state; hence, a measurement model for a \emph{single} measurement is sufficient.

\placeFig{2}

\paragraph*{Extent Model} For a given state $\xAll{k}$,  the target extent model specifies a single \emph{measurement source}  $\msourcer{k}[l]\in  \rvec{m}_k + \shapeSet{\xShape{k}}$  on the extended object (see \Fig{fig:meas_model_sequ}).\footnote{Set operations ``$+$`` and ``$\cdot$`` are meant element-wise.}
 In the following  \Sec{sec:rhm},  we introduce the  so-called \RHM\ for the target extent.

\paragraph*{Sensor Model}
For a given  measurement source  $\msourcer{k}[l]$, the sensor  model specifies the measurement $\meas{k}[l]$.
We focus on Cartesian point measurements corrupted with additive Gaussian noise according to
\begin{equation} \label{eqn:meas_eqn}
 \meas{k}[l]=  \msourcer{k}[l]+\mnoise{k}[l] \enspace, 
\end{equation}
where the noise term $\mnoise{k}[l]$  is zero-mean  white Gaussian noise with covariance matrix $\cov{v}[k][l]$.

\section{Dynamic Model}
In contrast to a point target, the temporal evolution of both the shape and kinematic parameters has to be modeled for an extended object.
 In this article,  we focus on  linear motion models according to
 \begin{equation} \label{eqn:dyn_eqn}
 \xAll{k+1}= \mat{A}_k  \xAll{k}+\snoise{k} \enspace, 
\end{equation}
where $\mat{A}_k$ is the system matrix  and $\snoise{k}$ is white Gaussian  system noise.
\chapter{Random Hypersurface Models}
 \label{sec:rhm}
In this section, a new target extent model called \RHM\ (\rhm) is introduced, which describes the location of a \emph{single}  measurement source in a  star-convex region.\footnote{The concept of an \rhm\ is much more general than presented here. The restriction to (two-dimensional) star-convex shapes is done for simplifying the following discussions.}

\begin{figure}
\center 
\includegraphics{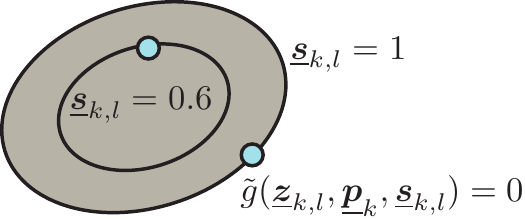}
\caption{\RHM\ (\rhm) for an ellipse\label{fig:rhm}.}
\end{figure}
\begin{Definition}[Star-Convexity]
 A set $\set{S}\subset \IR^2$ is \emph{star-convex} with respect to the origin iff $\tvect{0,0}\in \set{S}$ and  each line segment   from $\tvect{0,0}$ to any point in  $\set{S}$ is fully contained in $\set{S}$.
\end{Definition}

\section{Motivation: Implicit  Measurement Equation}\label{sec:rhmmotivation}
The objective is to  form  an equation that  relates the measurement source with the shape parameters.
When the measurement source $\msourcer{k}[l]$ lies on the boundary of the object, this is easy as
the boundary is a closed curve (in general a hypersurface) that can be described by an implicit equation in the form  $g(\msourcer{k}[l], \xShape{k})  =  0$.
However, the measurement sources may  also lie in the interior of boundary. In order to cover the interior, the basic idea is to scale the object boundary as described in  \Fig{fig:rhm}.
As the corresponding scaling factor $\rv{s}_{k,l} \in [0,1]$ for a measurement source   $\msourcer{k}[l]$ is unknown, we model it as a random variable and treat it as an additional noise term. By this means, we obtain the implicit relation 
\begin{equation}\label{eqn:implicitsourceeqn}
\tilde{g}(\msourcer{k}[l], \xShape{k}, \rv{s}_{k,l})  =  0
\end{equation}
which forms, together with  \Eq{eqn:meas_eqn},    an implicit measurement equation (note that $\tilde{g}(\msourcer{k}[l], \xShape{k},1)=g(\msourcer{k}[l], \xShape{k})$). 
In this sense,  modeling a two-dimensional region is reduced to modeling a curve by means of the scaling factor.

\placeFig{3}

\section{Definition}
According to the above motivation,  \rhms\ assume that the measurement source is  an element of a  scaled version of the  shape boundary, where the scaling factor is characterized by a particular  probability distribution, i.e., the scaling factor is modeled as a one-dimensional random variable.
 
\begin{Definition}[Random Hypersurface Model (RHM)]\label{def:rhm}
The measurement source $\msourcer{k}[l] \in \rvec{m}_k + \shapeSet{\xShape{k}}$ on an extended object with star-convex shape $\shapeSet{\xShape{k}} \subset \IR^2$ and center  $\rvec{m}_k \in \shapeSet{\xShape{k}}$ is generated according to an \rhm\ if  
   $$ \msourcer{k}[l] \in \rvec{m}_k+\rv{s}_{k,l}\cdot \boundSet{\xShape{k}}  \enspace ,$$ where 
$\boundSet{\xShape{k}}$ denotes the boundary of $\shapeSet{\xShape{k}}$    and     $\rv{s}_{k,l} \in [0, 1]$ is a  one-dimensional random variable.
\end{Definition}
 
 The restriction to star-convex shapes ensures that  $ \shapeSet{\xShape{k}} = \bigcup_{s \in [0,1]}\{ s \cdot \boundSet{\xShape{k}} \}$. In fact, scaling the object boundary corresponds to   a straight-line homotopy from the object center to the object boundary.
Note that all scaling factors  $\rv{s}_{k,l}$  are mutually  independent because measurements are assumed to be mutually independent.
 
\Def{def:rhm} does not specify where the measurement source $\msourcer{k}[l]$ lies on the scaled boundary.
In general, it is possible to consider it as an unknown fixed parameter  (\emph{functional model}) or to assume  it to be drawn from a probability distribution (\emph{structural model}). These two models are also widely-used  in curve fitting \cite{Chernov2009}.
For the structural model, an \rhm\ becomes a \emph{spatial distribution model} \cite{Gilholm2005,G2005}.

 \begin{Remark} The implicit representation motivated in \Sec{sec:rhmmotivation} is based on  the functional model. Probabilistic information about the location of a  measurement source on the boundary is not explicitely encoded in the implicit function.
  The shape boundary can be written as
 \begin{equation}
  \boundSet{\xShape{k}} =\{ \vec{z} \; | \; \vec{z} \in \IR^2 \text{ and }    g(\vec{z},\xShape{k})  = 0   \} \enspace,  
     \end{equation}
     and
     \begin{equation}
     \rv{s}_{k,l}\cdot  \boundSet{\xShape{k}} = \{ \vec{z} \; | \; \vec{z} \in \IR^2 \text{ and }    \tilde{g}(\vec{z},\xShape{k},\rv{s}_{k,l})  = 0   \} \enspace 
     \end{equation}
   is the scaled shape boundary.
  \end{Remark}

\section{Probability Distribution of the Scaling Factor} 
The probability distribution of the scaling factor depends on the distribution of the measurement sources on the extended object.
The following theorem  says how the scaling factor is distributed in case the measurement sources are uniformly distributed on the surface of a two-dimensional extended object.

 \begin{Theorem}\label{theo:uniform} 
If  the   measurement source $\rvec{z} \in \IR^2$ is uniformly distributed over the    star-convex region $\mathcal{S} \subset \IR^2 $ with center $\tvect{0,0}$, the corresponding squared scaling factor 
  is uniformly distributed on the interval $[0,1]$.
 \end{Theorem}
 \begin{proof}
The cumulative  distribution function $F(s)$ of $\rv{s}$ turns out  to be
\begin{equation}
 F(s) = \prob{\rv{s} \leq s}  =  \prob{\rvec{z} \in  s \cdot \mathcal{S} }   =    \frac{\text{Area}( s \cdot \mathcal{S})}{\text{Area}(\mathcal{S}) }\;=\;s^2   
\end{equation}
for $s \in [0,1]$. Additionally,    $F(s) = 0$ for $s< 0 $ and $F(s)=1$ for $s>1$. The cumulative distribution of  $\rv{u}:=\rv{s}^2$ given by $F(u)=\prob{\rv{s} \leq \sqrt{u}}$    is the cumulative  distribution function of the uniform distribution on $[0,1]$.
\end{proof}
According to \Theo{theo:uniform}, it is reasonable choose a uniform distribution for the squared scaling factor.

\section{General Procedure for Extended Object Tracking with RHMs}\label{sec:general}
In order to derive a state estimator for an extended object based on an \rhm , the following steps are to be performed:
\begin{itemize}
 \item A suitable shape and a shape parameterization have to be determined.
  \item The implicit equation \Eq{eqn:implicitsourceeqn} has to be formed. 
   \item  A  state estimator has to be derived based on the implicit measurement equation defined by  \Eq{eqn:implicitsourceeqn} and \Eq{eqn:meas_eqn}.
\end{itemize}
Based upon the above steps, we will derive  particular Gaussian estimators for ellipses and free-form star-convex shapes in this article.

\chapter{Formal Gaussian State Estimator  for Extended Objects}\label{sec:bayes_est}
In the following, the notation of a formal (Gaussian) Bayes filter for   the state $\xAll{k}$ based on the previously discussed models is introduced. Particular implementations  based on \rhms\ are presented in the next two sections.

We denote the probability density for the  parameter vector $\xAll{k}$   after the incorporation of  all measurements up to time step $k-1$ plus the measurements $\meas{k}[1],\ldots,\meas{k}[l]$ with         $f_{l}(\xAllDet{k})$ for $0\leq l \leq n_k$. In this article, we focus on Gaussian state estimators so that all probability densities are approximated with    Gaussians, i.e.,  $f_{l}(\xAllDet{k})\approx \Gauss{\xAllDet{k}}{\xAllMean{k}[l]}{\xAllCov{k}[l]}$, where $\xAllMean{k}[l]$ is the mean and $\xAllCov{k}[l]$ the covariance matrix.

\paragraph*{Time Update}
The time update  step predicts  $f_{n_{k-1}}(\xAllDet{k-1})$     to the next time step, i.e.,  it determines  $f_{0}( \xAllDet{k}  )$. 
As we focus on linear system models \Eq{eqn:dyn_eqn}, the time update can be performed with the Kalman filter formulas, see for example \cite{Bar-Shalom2002}.

\paragraph*{Measurement Update}
The prediction $f_{0}( \xAllDet{k})$ is  updated with the set of measurements   $\{ \meas{k}[l] \}_{l=1}^{n_k}  $   according to Bayes' rule.
Because the measurement generation process is assumed to be independent for consecutive measurements,  they can be incorporated recursively according to
$$f_{l}(\xAllDet{k})=\alpha_ {k,l}\cdot f(\meas{k}[l]| \xAllDet{k}  ) \cdot f_{l-1}( \xAllDet{k}  )    \text{  for  }  1\leq l \leq n_k    \enspace ,$$
where  $f(\meas{k}[l]| \xAllDet{k}) $ is a single measurement likelihood function and $\alpha_{k,l}$ is a normalization factor.
 
Note that the order of the measurements for a particular time step is irrelevant, because they are generated independent.
However, the processing order may matter if approximations are performed (see Remark~\ref{rem:order}).

\chapter{Elliptic Shapes}\label{sec:rhm_ellipse}

Elliptic shape approximations are highly relevant for real world applications as many targets are approximately elliptic.
Even in case of high measurement noise and few available measurements, an elliptic shape approximation  can be estimated.
In this section, an implicit measurement equation is derived for  elliptic shapes based on a  \rhm\ and subsequently a Gaussian state estimator is developed.

\section{Parameterization of an Ellipse}\label{sec:implict_rep_ellipse}
An ellipse is determined by its center $\rvec{m}_k \in \IR^2$ and a positive semi-definite  shape matrix $\mat{A}_k\in \IR^{2\times2}$.
Based on the  Cholesky decomposition $(\mat{A}_k)^{-1}=\mat{L}_k \mat{L}_k^T$, where
\begin{equation}\label{eqn:shape_para}
 \mat{L}_k:= \begin{bmatrix} \rv{a}_k & 0 \\   \rv{c}_k &  \rv{b}_k  \end{bmatrix} \enspace ,  
\end{equation}
a suitable vectorized parameterization  $ \xShape{k}[\text{el}]:= \begin{bmatrix}\rv{a}_k, \rv{b}_k, \rv{c}_k\end{bmatrix}^T $ of  $\mat{A}_k$    can be defined.
With $\xAll{k}[\text{el}]:=\tvect{ \rvec{m}^T_k, (\rvec{x}^*_k)^T, (\xShape{k}[\text{el}])^T }$,  the implicit shape function becomes 
$$g^{\text{el}}(\msourcer{k}[l],\xAll{k}[\text{el}]):= (\msourcer{k}[l] - \rvec{m}_k)^T \cdot  \mat{A}_k^{-1}      \cdot  (\msourcer{k}[l]- \rvec{m}_k) -1\enspace .  $$

\section{Implicite Measurement Equation}
Thanks to the chosen representation of an ellipse, the  implicit function for the scaled version of $\boundSet{\xShape{k}[\text{el}]}$ with scaling factor $\rv{s}_{k,l}$  turns out to be
\begin{equation}\label{eqn:meas_eqn_elliptic1}
\tilde{g}^{\text{el}}(\msourcer{k}[l],\xAll{k}[\text{el}], \rv{s}_{k,l}):= (\msourcer{k}[l] - \rvec{m}_k)^T \cdot  \mat{A}_k^{-1}      \cdot  (\msourcer{k}[l]- \rvec{m}_k) -\rv{s}_{k,l}^2\enspace .  
\end{equation}
 Equation \Eq{eqn:meas_eqn_elliptic1} and  \Eq{eqn:meas_eqn}  specify together an implicit measurement equation, which in this case coincides  with the problem of fitting an ellipse to noisy data points plus the additional random scaling factor.

\section{Gaussian State Estimator}\label{ssec:ellipse_bayesest}

Approaches based on the Kalman filter for state estimation with implicit measurement equations such as \Eq{eqn:meas_eqn_elliptic1} and  \Eq{eqn:meas_eqn} are well-known in literature. Typically, the implicit measurement equation is linearized around the measurement and state in order to render it explicit, see for example \cite{Soatto1996}. Here, we propose to perform algebraic reformulations  followed by a statistical linearization around the measurement source as described in the following. When \Eq{eqn:meas_eqn_elliptic1} would be linear, we could rewrite the problem directly as an explicit measurement equation by plugging \Eq{eqn:meas_eqn} into  \Eq{eqn:meas_eqn_elliptic1}. As  \Eq{eqn:meas_eqn_elliptic1} is nonlinear,   an exact reformulation is not possible. Nevertheless this reformulation can be performed approximately. For this purpose, the first step is to plug   \Eq{eqn:meas_eqn} into \Eq{eqn:meas_eqn_elliptic1}, i.e.,
\begin{equation}\label{eqn:meas_noise_approx_elliptic}
 \tilde{g}^{\text{el}}( \meas{k}[l], \xAll{k}[\text{el}], \rv{s}_{k,l})  =  \tilde{g}(\msourcer{k}[l]+\mnoise{k}[l],\xShape{k}[\text{el}],\rv{s}_{k,l}) 
 \enspace .
\end{equation}
After some minor simplifications  of \Eq{eqn:meas_noise_approx_elliptic} and exploiting \Eq{eqn:meas_eqn_elliptic1}, the following measurement equation  
\begin{equation}
h^{\text{el}}(\xAll{k}[\text{el}],\mnoise{k}[l], \rv{s}_{k,l},\meas{k}[l] ) =  0  \label{eqn:meas_approx_elliptic} \enspace 
\end{equation}
with a  pseudo-measurement $0$ is obtained, where 
\begin{equation}
h^{\text{el}}(\xAll{k}[\text{el}],\mnoise{k}[l], \rv{s}_{k,l},\meas{k}[l] ):= \tilde{g}^{\text{el}}(\meas{k}[l],\xShape{k},\rv{s}_{k,l})- 2(\msourcer{k}[l]-\rvec{m}_k)^T \mat{A}_k^{-1} \mnoise{k}[l]  -    \mnoise{k}[l]^T \mat{A}_k^{-1} \mnoise{k}[l]   -  \rv{s}_{k,l}^2 \enspace
\end{equation}
maps the state $\xAll{k}[\text{el}]$,  the measurement noise $\mnoise{k}[l]$, the scaling factor $\rv{s}_{k,l}$, and the measurement $\meas{k}[l]$ to the pseudo-measurement.

However,   the unknown measurement source $\msourcer{k}[l]$ still occurs in \Eq{eqn:meas_approx_elliptic}.  The basic idea here is to substitute $\msourcer{k}[l]-\rvec{m}_k$   in \Eq{eqn:meas_approx_elliptic} with a proper point estimate. The easiest way to obtain a point estimate for $\msourcer{k}[l]-\rvec{m}_k$ is to consider the  ellipse specified by the mean of the previous estimate, i.e.,  $\xShapeMean{k}[l-1]$ and use  the point with the smallest distance from the conic to the measurement $\meas{k}[l]$ as a point estimate for obtaining  a point estimat for  $\msourcer{k}[l]-\rvec{m}_k$.

The measurement equation \Eq{eqn:meas_approx_elliptic} can  directly be used within Gaussian state estimators such as the Unsented Kalman Filter (UKF) \cite{Julier_UnscentedFiltering} or analytic moment calculation \cite{Fusion11_Baum}, which essentially performs a statistical linearization of  \Eq{eqn:meas_approx_elliptic}. More precisely,  the  joint density of the state and the predicted pseudo-measurement is approximated with a Gaussian. Then, the Kalman filtering \cite{Bar-Shalom2002} formulas can be used for calculating the updated estimate, i.e., 
given the  previous estimate $f_{l-1}(\xAllDet{k})= \Gauss{\xAllDet{k}}{\xAllMean{k}[l-1]}{\xAllCov{k}[l-1]}$, the updated estimate 
$f_{l}(\xAllDet{k})= \Gauss{\xAllDet{k}}{\xAllMean{k}[l]}{\xAllCov{k}[l]}$
 with the measurement $\meas{k}[l]$ results from the Kalman filter
equations\footnote{For sake of clarity, the index ``$\text{el}$''  is omitted for the mean and covariance matrices of the estimates.}
\begin{eqnarray}\label{eqn:update_ellipse}
 \xAllMean{k}[l]  &=&    \xAllMean{k}[l-1]    + \cov{xh}[k]  (\cov{hh}[k] )^{-1} \left(0-\mean{h}[k]  \right) \enspace ,\\
\xAllCov{k}[l] &=&   \xAllCov{k}[l-1] - \cov{xh}[k] (\cov{hh}[k] )^{-1} \cov{hx}[k] \label{eqn:update_ellipse2} \enspace ,
\end{eqnarray}
where   $0$ is the predicted pseudo-measurement, $\cov{xh}[k]$ is the  covariance between the pseudo-measurement and   the state,  and $\cov{hh}[k]$ is the variance of the predicted pseudo-measurement.
In this manner, a statistical linearization around the measurement source is performed.
 At this point, it is important to note that $\mean{h}[k]$ and  $\cov{xh}[k]$ do  not depend on the unknown measurement source $\msourcer{k}[l]$ and, hence, the error made due to substituting it with a point estimate is rather negligible.

\begin{Remark}\label{rem:order}
 Due to the nonlinear measurement model and the performed approximations, the order of the measurement processing matters. However, we observed that the differences arising from different processing orders is rather negligible.
  Besides, it is possible to perform a batch processing of all measurements by means of stacking the single measurement functions \Eq{eqn:meas_approx_elliptic}. By this means, the estimation accuracy can be slightly increased.
\end{Remark}
\begin{Remark}\label{rem:approx_error}
Due to the statistical linearization of the implicit measurement equation, approximation errors are introduced. As a consequence, the resulting estimation quality highly  depends on the
parameterization of the ellipse and of the particular form of $\tilde{g}^{\text{el}}(\meas{k}[l], \xAll{k}[\text{el}],   \rv{s}_{k,l})$.
For example, we observed that multipliying both sides of  \Eq{eqn:meas_noise_approx_elliptic} with $1/\Tr{\mat{L}_k  \mat{L}_k^T}$ improves the estimation quality.
 \end{Remark}
\begin{Remark}
Note that the  actual likelihood function  $f(\meas{k}[l]| \xAllDet{k}[\text{el}]) $  can be derived from  the measurement equation~\Eq{eqn:meas_starconvex4}, however, it is not required explicitely when performing statistical linearization.
\end{Remark}

\chapter{Star-Convex Shapes}\label{sec:rhm_starconvex}
When the measurement noise is low compared to the target extent, it may be 
possible to extract more detailed shape information than an ellipse from the measurements. 
For this purpose, an \rhm\ for estimating and tracking the parameters of a star-convex  shape approximation is presented in the following. A detailed target shape approximation is of high value for many higher-level problems such as target classification,
track management, and  sensor management. Last but not least, a more detailed shape estimate results in a better estimation quality for the kinematic state of the target as it is a more precise model of the reality.

\section{Parametrization of Star-Convex Shapes}

A star-convex shape $\shapeSet{\xShape{k}[\text{sc}]}$ can be represented in \emph{parametric form}  with the  help of   a   radius function
\begin{figure*}
 \centering
 \subfloat[Star-convex shape.]{
 \psfragfig*[width=4cm]{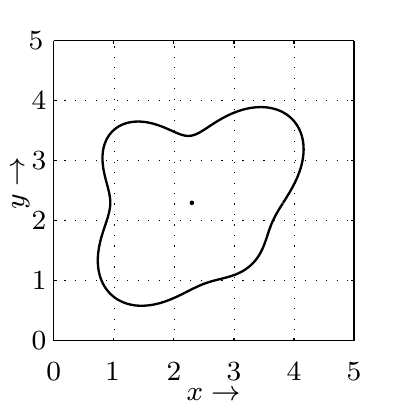}
 \label{fig:polar1}
}  
\subfloat[Radius function.]{
 \psfragfig[width=4cm]{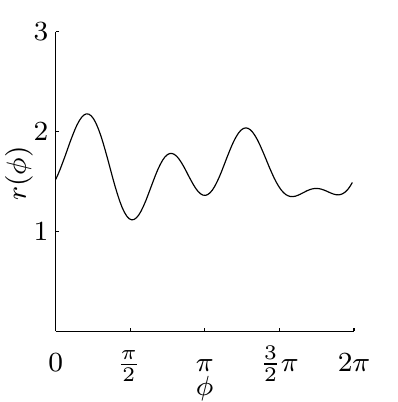}
 \label{fig:polar2}
}  
\caption{Representation of a star-convex shape with a radius function.\label{fig:polar}}
\end{figure*}
 $r(\xShape{k}[\text{sc}],\phi)$, which gives the distance from the object center to  a contour point depending on the angle $\phi$ and a parameter vector $\xShape{k}[\text{sc}]$ (see \Fig{fig:polar}).       
A suitable (finite dimensional) parameterization $\xShape{k}[\text{sc}]$ of a radius function is given by  the first $N_f$ Fourier coefficients that define a  Fourier series expansion \cite{Uensalan2001,Zhang2005}, i.e.,
\begin{equation*}\label{eqn:fourier_linear}
r(\xShape{k}[\text{sc}],\phi)=  \tfrac{1}{2}\rv{a}_k^{(0)} +\sum_{j=1\ldots N_F }\(\rv{a}_k^{(j)} \cos(j\phi) + \rv{b}_k^{(j)} \sin(j\phi) \) =\mat{R}(\phi) \cdot \xShape{k}[\text{sc}] \enspace , 
\end{equation*}
where
\begin{eqnarray*}
\mat{R}(\phi)&=& [\tfrac{1}{2}, \cos(\phi), \sin(\phi), \ldots , \cos(N_F\phi), \sin(N_F\phi)  ] \enspace , \text{ and }\\ 
\xShape{k}[\text{sc}]&=& \tvect{\rv{a}_k^{(0)}, \rv{a}_k^{(1)},  \rv{b}_k^{(1)}, \ldots \rv{a}_k^{(N_F)},\rv{b}_k^{(N_F)} } \enspace . \end{eqnarray*}
Fourier coefficients  with small indices  encode information about the coarse features of the shape and Fourier coefficients with larger indices  encode finer details.

\placeFig{4}

\section{Implicit Measurement Equation} 
With $\xAll{k}[\text{sc}]:=\tvect{ \rvec{m}^T_k, (\rvec{x}^*_k)^T, (\xShape{k}[\text{sc}])^T }$ and the implicit representation of star-convex curves \cite{Uensalan2001}, we obtain 
\begin{equation}\label{eqn:sc_impl_ms1}
g^{\text{sc}}(\msourcer{k}[l],\xAll{k}[\text{sc}])=  ||\boldsymbol{\underline{m}} -  \msourcer{k}[l]||^2-  r( \xShape{k}[\text{sc}] , \measuredangle(\rv{m} - \msourcer{k}[l]  )  )^2 \enspace,  
\end{equation}
 where $\measuredangle(\rv{m} - \msourcer{k}[l]  )$      denotes the angle between $\rv{m} - \msourcer{k}[l]$ and the $x$-axis.
The scaled version of the shape boundary is specified by
\begin{equation}\label{eqn:sc_impl_ms2}
\tilde{g}^{\text{sc}}(\msourcer{k}[l],\xAll{k}[\text{sc}],  \rv{s}_{k,l})=  ||\boldsymbol{\underline{m}} -  \msourcer{k}[l]||^2- \boldsymbol{s}_{k,l}^2 \cdot r( \xShape{k}[\text{sc}] , \measuredangle(\rv{m} - \msourcer{k}[l]  )  )^2 \enspace,  
\end{equation}
Again,  \Eq{eqn:sc_impl_ms2} specifies together with \Eq{eqn:meas_eqn} an implicit measurement equation.

\section{Bayesian State Estimator}\label{ssec:starconvex_bayesest}
A measurement equation can be derived  by  plugging \Eq{eqn:meas_eqn} into  \Eq{eqn:sc_impl_ms2}
\begin{equation}\label{eqn:meas_noise_approx_ellipticSDASdasd}
 \tilde{g}^{\text{sc}}( \meas{k}[l] , \xAll{k}[\text{sc}],\rv{s}_{k,l})  =  \tilde{g}( \msourcer{k}[l]+\mnoise{k}[l], \xShape{k}[\text{sc}],\rv{s}_{k,l})  \enspace .
\end{equation}
In order to avoid the treatment of uncertain angles, we propose to replace the angles in  
occurrences $r( \xShape{k}[\text{sc}] , \cdot )$ with a point estimate $\hat{\phi}_{k,l}$, i.e.,  we assume $r( \xShape{k}[\text{sc}] , \measuredangle(\rv{m} - \meas{k}[l] )) \approx         r( \xShape{k}[\text{sc}] , \measuredangle(\rv{m} - (\msourcer{k}[l]+\mnoise{k}[l])  ) \approx  r( \xShape{k}[\text{sc}] , \hat{\phi}_{k,l})$.

\begin{Remark}
A proper point estimate is given by the most likely angle $\phi_{k,l}$.
In case of isotropic measurement noise,  this  point estimate  $\phi_{k,l}$ is  given by the angle between the vector from the current shape center estimate $\vec{\mu}^m_{k,l-1}$   to the measurement $\meas{k}[l] $ and the $x$-axis, i.e., $\hat{\phi}_{k,l} := \measuredangle \big( \meas{k}[l] - \vec{\mu}^m_{k,l-1}  \big)$.  
\end{Remark}

Based on the point estimate and  $\boldsymbol{\underline{m}_k} -  \msourcer{k}[l] \approx \rv{s}_{k,l}\cdot \mat{R}(\hat{\phi}_{k,l}) \cdot \xShape{k}[\text{sc}] \cdot  \vec{e}(\hat{\phi}_{k,l}) $, where $\vec{e}(\hat{\phi}_{k,l}):= \tvect{\cos{\hat{\phi}_{k,l}},\sin{\hat{\phi}_{k,l}}}$, \Eq{eqn:meas_noise_approx_ellipticSDASdasd} can be simplified to the following measurement equation 
\begin{equation}
h^{\text{sc}}(\xAll{k},\mnoise{k}[l], \rv{s}_{k,l},\meas{k}[l] ) =  0  \label{eqn:meas_starconvex4} \enspace 
\end{equation}
with
\begin{equation}
h^{\text{sc}}(\xAll{k}[{\text{sc}}],\mnoise{k}[l], \rv{s}_{k,l},\meas{k}[l] )  :=\rv{s}_{k,l}^2 \cdot  ||\mat{R}(\hat{\phi}_{k,l})\cdot  \xShape{k}[\text{sc}]||^2 +   
    2 \rv{s}_{k,l}\mat{R}(\hat{\phi}_{k,l})\xShape{k}[\text{sc}]  \vec{e}(\hat{\phi}_{k,l})^T \mnoise{k}[l]              +  ||\mnoise{k}[l]||^2-      
  || \meas{k}[l] - \rvec{m}_k  ||^2 \enspace ,   \notag
\end{equation}
which maps the state $\xAll{k}[{\text{sc}}]$,  the measurement noise $\mnoise{k}[l]$, the scaling factor $\rv{s}_{k,l}$, and the measurement $\meas{k}[l]$ to a pseudo-measurement $0$.

\begin{Remark}
 An alternative derivation of \Eq{eqn:meas_starconvex4} based on the explicit representation of the shape with the  radius function can be found in \cite{Fusion11_Baum}.
\end{Remark}

For a given  density $f_{l-1}(\xAllDet{k})= \Gauss{\xAllDet{k}}{\xAllMean{k}[l-1]}{\xAllCov{k}[l-1]}$, the posterior density  $f_{l}(\xAllDet{k})= \Gauss{\xAllDet{k}}{\xAllMean{k}[l]}{\xAllCov{k}[l]}$   having received the measurement  $\meas{k}[l]$ can be calculated with a Gaussian state estimator such as the UKF \cite{Julier_UnscentedFiltering}  or analytic moment calculation \cite{Fusion11_Baum} for a    closed-form measurement update. See also  \Eq{eqn:update_ellipse} and \Eq{eqn:update_ellipse2} in this context.
Just as for ellipses, the order of the measurement processing matters (see the discussion in \Rem{rem:order}).

\chapter{Evaluation}\label{ssec:eval_ellipse}\label{ssec:eval_starconvex}
\rhms\ for elliptic and star-convex shapes are evaluated by means of both stationary and moving extended objects.\footnote{Source code for \rhms\ is available at  {\verbBox{http://www.cloudrunner.eu}}.}
For both elliptic shapes and star-convex shapes the UKF \cite{Julier_UnscentedFiltering} is used for performing the measurement update.
For elliptic shapes, the \emph{squared} scaling factor is modeled as a Gaussian distribution with mean 0.5 and variance $1/12$ (i.e., the first two moments of a uniform distribution).
For star-convex shapes, the scaling factor is modeled as Gaussian distribution with mean 0.7 and variance $0.06$.

\section{Stationary Extended Target}\label{ssec:eval_ellipse_static}

\begin{figure*}
\center
\subfloat[Target 1: Elliptic shape.]{
  \psfragfig[width=4cm]{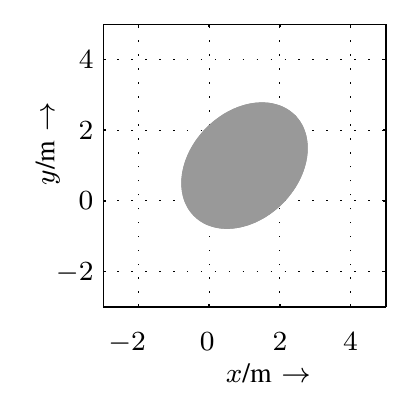}
 \label{fig:sim_static_targets1} \label{fig:sim_ellipse_target1}
}  
\subfloat[Target 2: Aircraft-like shape.]{
\psfragfig[width=4cm]{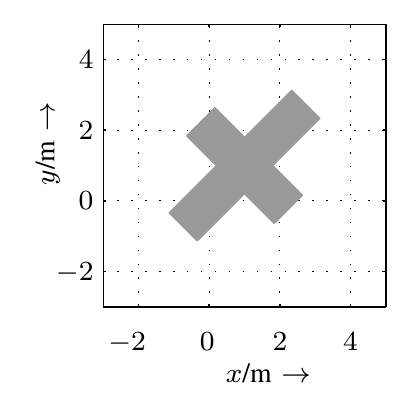}
 \label{fig:sim_static_targets1}
}  
\subfloat[Target 3: Group target.]{
   \psfragfig[width=4cm]{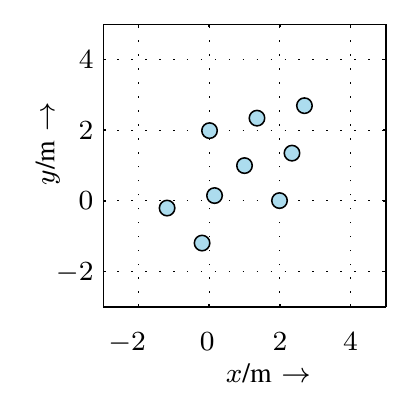}
 \label{fig:sim_static_targets2} \label{fig:sim_ellipse_target2}
}  
 \caption{Targets used for the evaluation\label{fig:sim_ellipse_targets}. \label{fig:sim_static_targets}} 
\end{figure*}

\begin{figure*}
\subfloat[Low noise level.]{
 \hspace{-0.5cm}
\psfragfig[width=4.2cm]{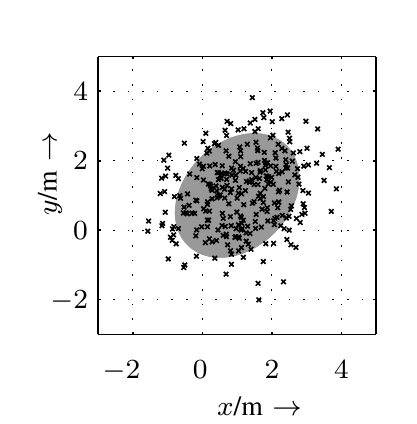}
 \hspace{-0.2cm}
\psfragfig[width=4.2cm]{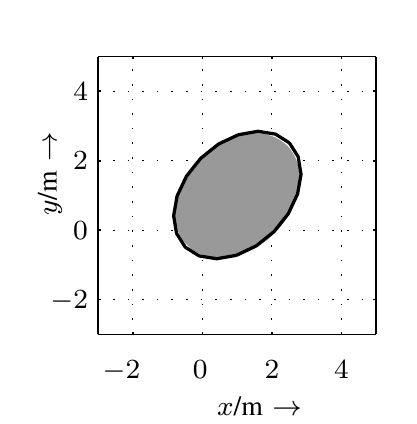}
  \psfragfig[width=4.2cm]{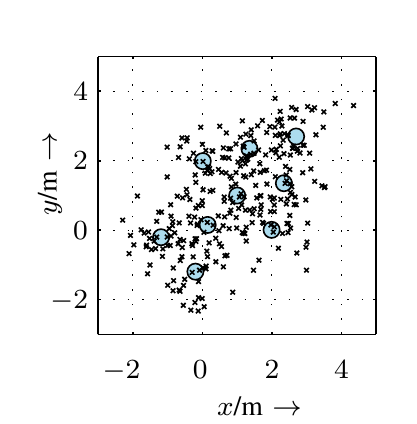}
   \hspace{-0.2cm}
 \psfragfig[width=4.2cm]{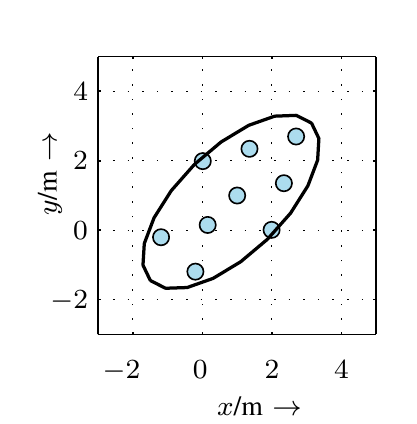}
}  

\subfloat[Medium noise level.]{
  \hspace{-0.5cm}
 \psfragfig[width=4.2cm]{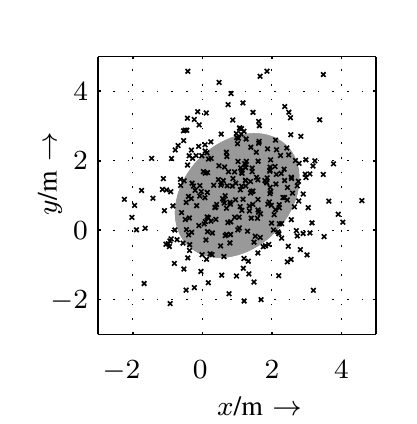}
 \hspace{-0.2cm}
 \psfragfig[width=4.2cm]{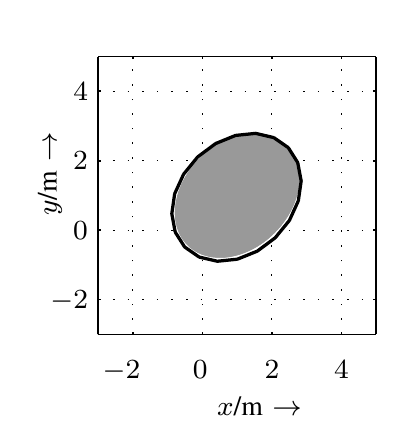}
   \psfragfig[width=4.2cm]{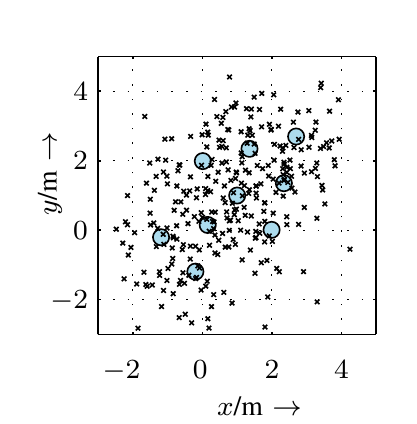}
\hspace{-0.2cm}
   \psfragfig[width=4.2cm]{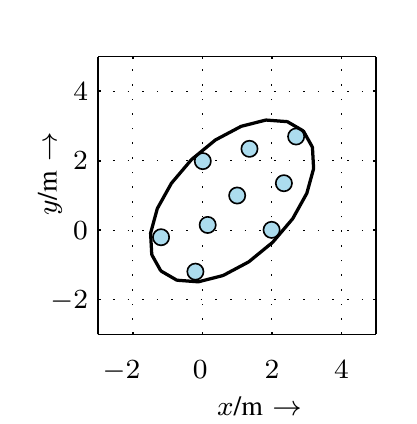}
}  
 
\subfloat[High noise level.]{
  \hspace{-0.5cm}
 \psfragfig[width=4.2cm]{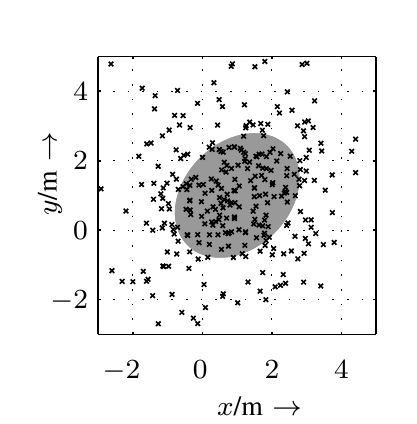}
 \hspace{-0.2cm}
   \psfragfig[width=4.2cm]{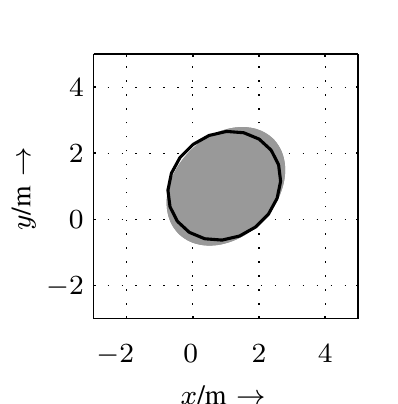}
   \psfragfig[width=4.2cm]{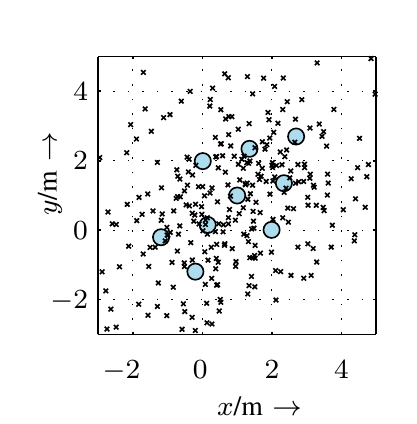}
 \hspace{-0.2cm}
   \psfragfig[width=4.2cm]{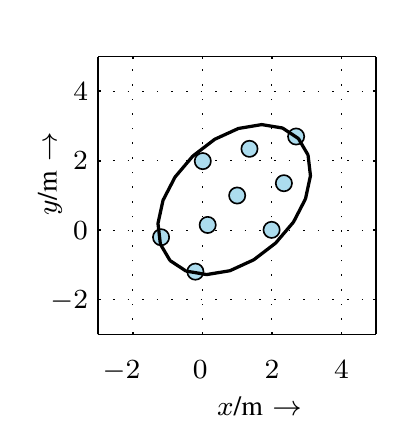}
}  
\caption{Simulation results: Example measurements for a particular run and point estimates for the shape averaged over $20$ runs\label{fig:sim_ellipse_results}. Simulations are performed with low measurement noise level $\cov{v}[k][1]=\diag(0.6^2,0.6^2) $, medium measurement noise level $\cov{v}[k][1]=\diag(1,1)$, and high measurement noise level $\cov{v}[k][1]=\diag(1.4^2,1.4^2)$.
}

 \end{figure*}

\begin{figure*}
\subfloat[Low  noise level.]{
 \hspace{-0.2cm}
 \psfragfig[width=4cm]{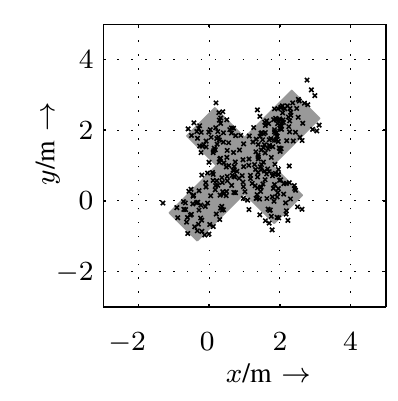}
  \hspace{-0.2cm}
 \psfragfig[width=4cm]{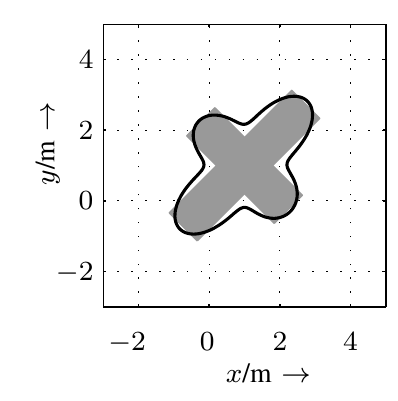}
\psfragfig[width=4cm]{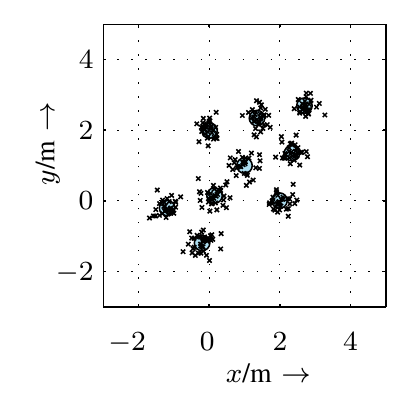}
  \hspace{-0.2cm}
 \psfragfig[width=4cm]{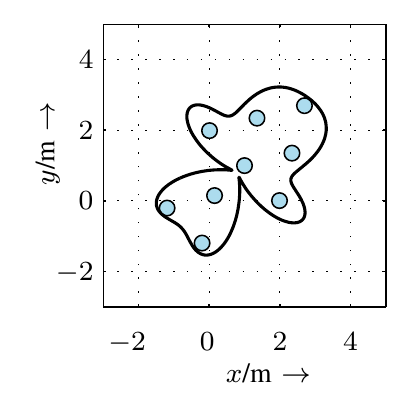}
   
}  

\subfloat[Medium noise level.]{ 
\hspace{-0.2cm}
\psfragfig[width=4cm]{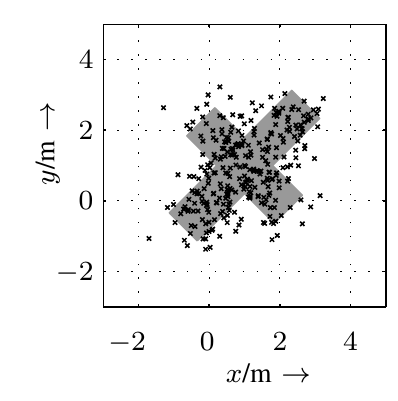}
\hspace{-0.2cm}
\psfragfig[width=4cm]{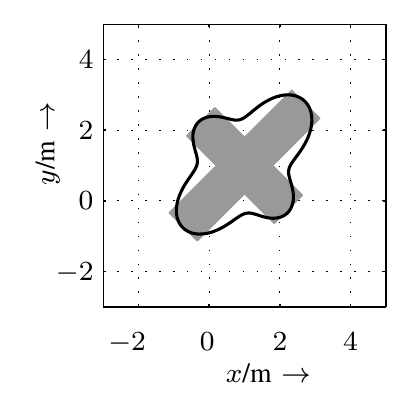}
\psfragfig[width=4cm]{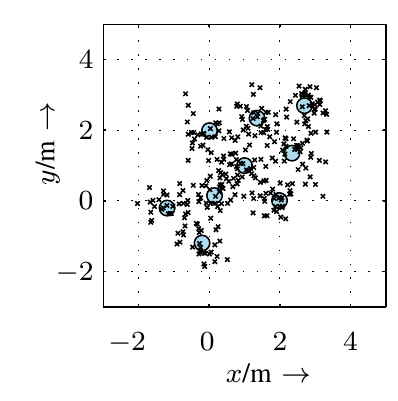}
\hspace{-0.2cm}
\psfragfig[width=4cm]{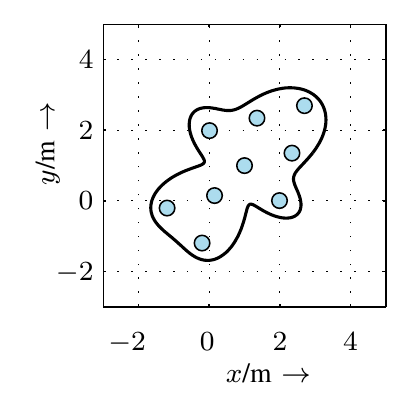}
}  

\subfloat[High noise level.]{
\hspace{-0.2cm}
\psfragfig[width=4cm]{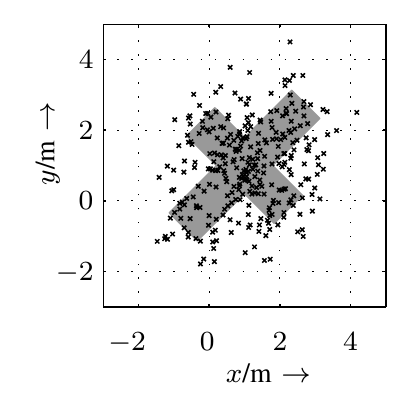}
\hspace{-0.2cm}
\psfragfig[width=4cm]{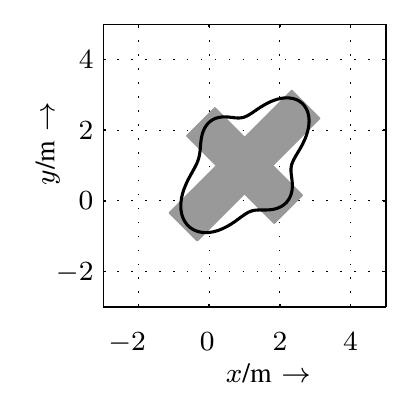}
\psfragfig[width=4cm]{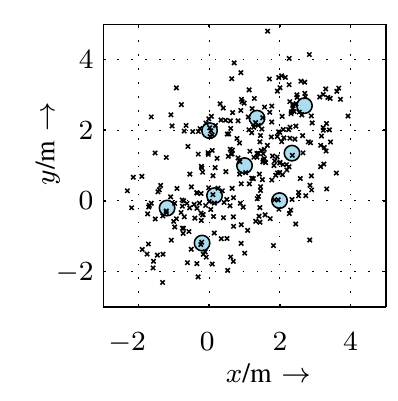}
\hspace{-0.2cm}
 \psfragfig[width=4cm]{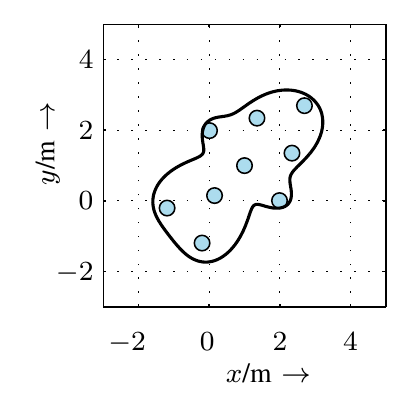}
}  
\caption{Simulation results: Example measurements for a particular run  and point estimates for the shape averaged over $20$ runs\label{fig:sim_starconv_results}. Simulations are performed with low measurement noise level $\cov{v}[k][1]=\diag(0.3^2,0.3^2)$, medium measurement noise level $\cov{v}[k][1]=\diag(0.4^2,0.4^2)$, and high measurement noise level $\cov{v}[k][1]=\diag(0.6^2,0.6^2)$.
}
 \end{figure*}

\begin{figure*}
\subfloat[First measurement.]{
\hspace{-0.3cm}
\psfragfig[width=4cm]{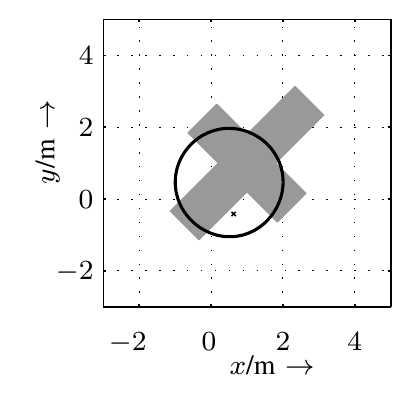}
 }
 \subfloat[10 measurements.]{
  \hspace{-0.2cm}
 \psfragfig[width=4cm]{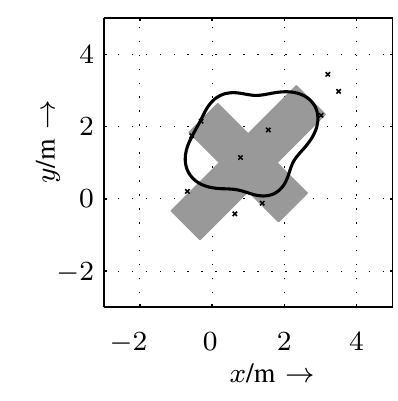}
 }
 \subfloat[30 measurements.]{
  \hspace{-0.2cm}
 \psfragfig[width=4cm]{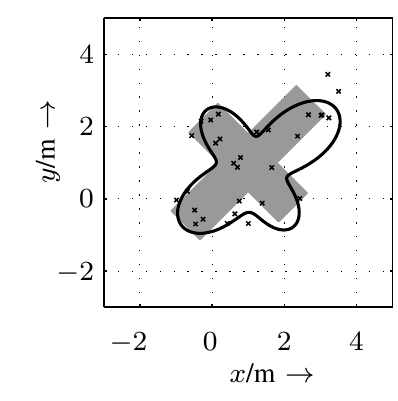}
 }
 \subfloat[50 measurements.]{
  \hspace{-0.2cm}
 \psfragfig[width=4cm]{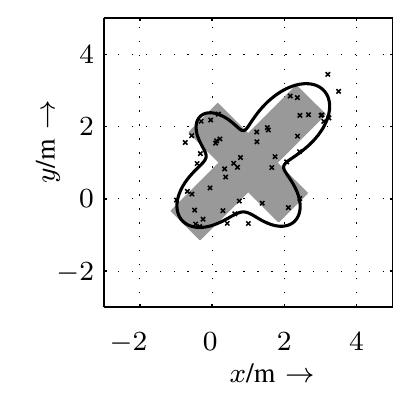}
 }
 \caption{Example run demonstrating the sequential incorporation of measurements (Noise level: $\cov{v}[k][1]=\diag(0.3,0.3)$). \label{fig:sim_ellipse_sequential}}
\end{figure*}

In the first scenario, we consider an extended object with  fixed position and shape.
From the target,  $300$ measurements are received sequentially, i.e., a single measurement per time step~ ($n_k=1$).
Simulations are performed with  the three different target types depicted in \Fig{fig:sim_ellipse_targets}, where 
measurement sources are drawn uniformly from the target surface/group members.

The parameters of the ellipse are a priori set to a Gaussian with mean $\tvect{0.5, 0.5, 1.6, 1.6, 0.6}$ and covariance matrix $\diag(3, 3, 0.5, 0.5, 0.5)$, i.e., an uncertain circle with  radius $1.2$ and center  $\tvect{0.5, 0.5}$.
For the star-convex shape approximation,
 $15$ Fourier coefficients are used, where the shape parameters are a priori set to a Gaussian with mean $\tvect{0.5, 0.5, 3, 0, \ldots, 0}$ and covariance matrix $\diag(0.7, 0.7, 0.1, 0.1, \ldots ,0.1)$, i.e., an uncertain circle with  radius $1.5$ and center  $\tvect{0.5, 0.5}$.

The estimation results after the $300$ measurements are depicted in \Fig{fig:sim_ellipse_results} and \Fig{fig:sim_starconv_results} for different measurement noise levels.  
The shape estimates are averaged over $20$ Monte-Carlo runs. 
In order to  illustrate the magnitude of the measurement noise, the measurements of a particular run are also plotted. It is important to note that this is just done for visualization as the estimator incorporates the measurements  recursively.
 \Fig{fig:sim_ellipse_sequential}  depicts a single example run for star-convex shapes  in order to show the evolution of the shape estimates with an increasing number of measurements.

\placeFig{5}

\placeFig{6}

\placeFig{7}

 \placeFig{8}

\section{Moving Extended Object}\label{ssec:eval_ellipse_dyn}
\begin{figure}
 \centering
 \subfloat[Target shape.]{
\psfragfig[width=4cm]{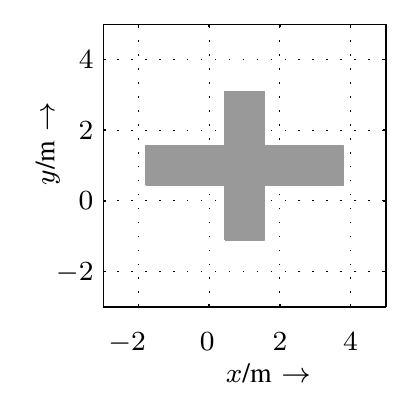}
 \label{fig:sim_dyn_tar}
}
\subfloat[Trajectory of the target.]{
  \psfragfig[width=4.5cm]{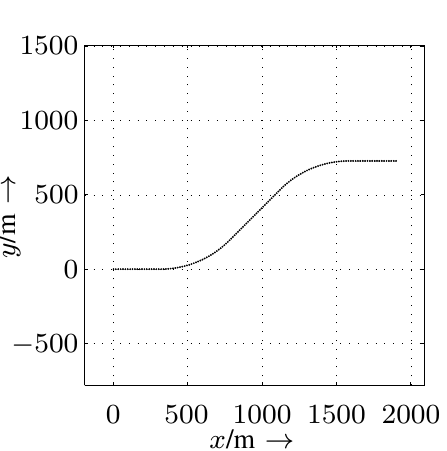}
\label{fig:sim_dyn_tra}
} 
\caption{Extended object and its trajectory.}
\end{figure}

In the second scenario, the aircraft-shaped target  shown in \Fig{fig:sim_dyn_tar}  moves along the trajectory depicted in \Fig{fig:sim_dyn_tra}.
The measurement sources are drawn uniformly from the target surface.
The magnitude of the measurement noise varies from measurement to measurement in order to simulate different sensors or different target-to-sensor geometries.
The covariance matrix of the measurement noise is 
 $\cov{v}[k][1]=\diag(0.2^2,0.2^2)$ with probability $0.75$ and $\cov{v}[k][1]=\diag(0.4^2,0.4^2)$ with probability $0.25$. 
The number of measurements received per time instant is given by  $n_k = n^*_k  + 1 $, where $n^*_k$  is a Poisson distributed random variable  with mean $4$ for the ellipse and mean $7$ for the star-convex shape.

We employ a constant velocity model for the temporal evolution of the target center \cite{Bar-Shalom2002} and
a random walk model for the shape parameters.
Hence, the state vector to be tracked is $\xAll{k}=\tvect{ \rvec{m}_k , \rvec{m}^v_k ,    {(\xShape{k}[\text{sc}])^T} }$, where  $\rvec{m}_k$ is the center,  $ \rvec{m}^v_k $ is the velocity vector, and $\xShape{k}[\text{sc}]$ are the shape parameters.
As the center is assumed to evolve according to a constant velocity model, the  system matrix  in \Eq{eqn:dyn_eqn} is 
 $\mat{A}_k= \diag(\mat{A}^{cv}_k, \mat{I}_d  )$, where  $\mat{A}^{cv}_k=  \vect{\mat{I}_{2} & T \mat{I}_{2} \\   0 &\mat{I}_{2}}$  with $T=1$ and $\mat{I}_d$ is the identity matrix with dimension $d=5$ for ellipses and $d=11$ for star-convex shapes.
The system noise is zero-mean Gaussian noise with covariance matrix
$\mat{C}_k^{w}=\diag( q_1\cdot \mat{I}_{d}, \mat{C}_k^{cv})$  with 
$\mat{C}_k^{cv}= q_2 \vect{\frac{T^3}{3}\mat{I}_{2} &  \frac{T^2}{2}\mat{I}_{2}\\     \frac{T^2}{2}\mat{I}_{2} & T\mat{I}_{2}  }$.
For the ellipse, we have $q_1=0.0015$ and $q_2=0.005$. For the star-convex shape  $q_1=0.0001$ and $q_2=0.003$.

\begin{figure*}
\subfloat[Example measurements.]{
 \hspace{-0.2cm}
 \psfragfig[width=4cm]{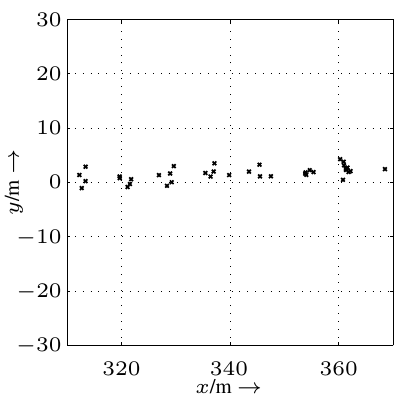}
}  
\subfloat[Shape estimates.]{
 \hspace{-0.2cm}
 \psfragfig[width=4cm]{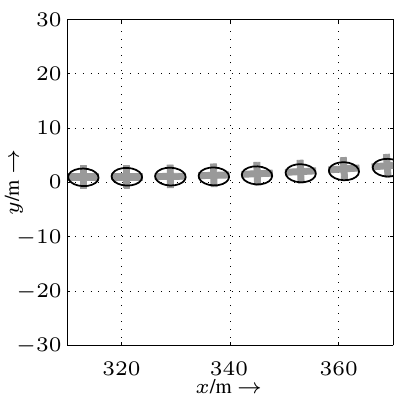}
}  
\subfloat[Example measurements. ]{
\hspace{-0.2cm}
\psfragfig[width=4cm]{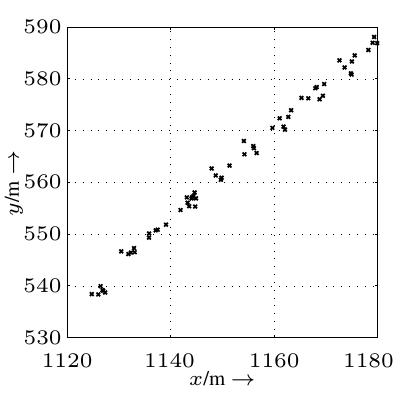}
}  
\subfloat[Shape estimates.]{
\hspace{-0.2cm}
\psfragfig[width=4cm]{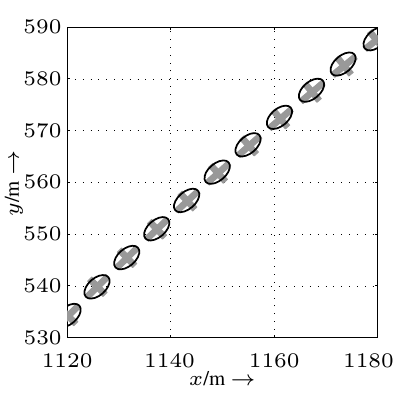}
}  
 
\caption{Tracking an extended object with an \rhm\ for ellipses: Example measurements from a particular run and average shape estimates. \label{fig:sim_ellipse_dyn}}
\end{figure*}

\begin{figure*}
\subfloat[Example measurements.]{
\hspace{-0.2cm}
\psfragfig[width=4cm]{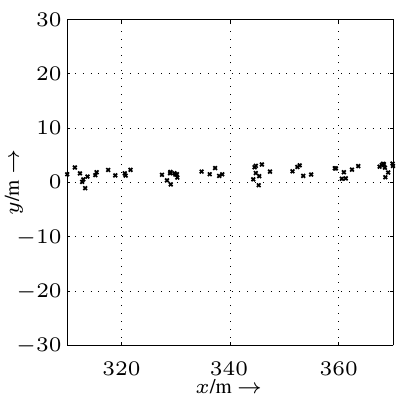}
}  
\subfloat[Shape estimates.]{
\hspace{-0.2cm}
\psfragfig[width=4cm]{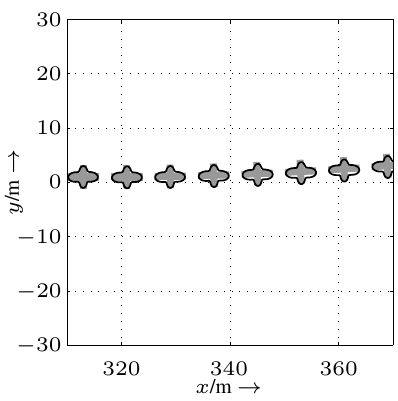}
}  
\subfloat[Example measurements. ]{
 \hspace{-0.2cm}
 \psfragfig[width=4cm]{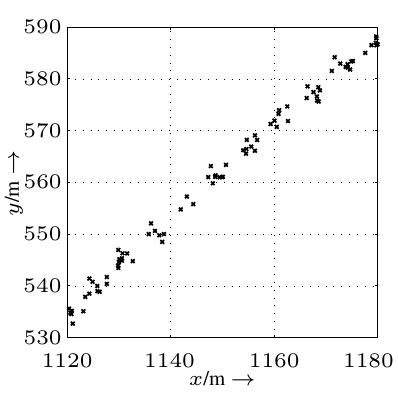}
}  
\subfloat[Shape estimates.]{
\hspace{-0.2cm}
\psfragfig[width=4cm]{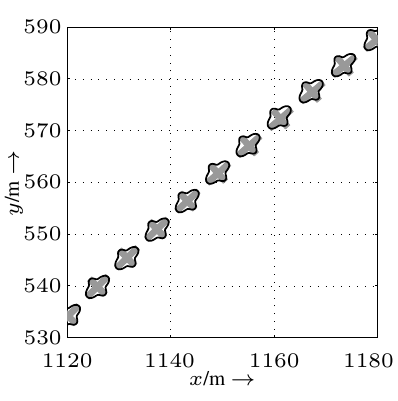}
}  
\caption{Tracking an extended object with an \rhm\ for star-convex shapes: Example measurements from a particular run and average shape estimates. \label{fig:sim_starconvex_dyn}}
\end{figure*}

The estimated shapes (averaged over $20$ time steps) are depicted in  \Fig{fig:sim_ellipse_dyn} and  \Fig{fig:sim_starconvex_dyn} for two snippets of the trajectory. 
The results show that the shape of the extended object is tracked precisely, even when the shape changes its orientation. 

Note that  in the simulations with elliptic  shapes the number of measurements per time step is lower than for star-convex shapes.
A star-convex shape approximation can only be extracted well in case the measurements carry enough information, i.e., enough measurements with rather low noise are available.

The example measurements in \Fig{fig:sim_ellipse_dyn} and  \Fig{fig:sim_starconvex_dyn}  also emphasize that na\"\i ve  approaches for estimating a shape would be bound to fail,
e.g., directly computing  an enclosing shape of the measurements is infeasible because the measurements are noisy  and only a couple of measurements are available  per time step.

\placeFig{9}

\placeFig{10}

\placeFig{11}

\chapter{Conclusions and Future Work}\label{sec:conclusions}
This article considered the problem of estimating a shape approximation of an extended object, which gives rise to several measurements from different spatially distributed measurement sources. 
For this purpose, a novel approach for modeling  extended objects called \RHM\ (\rhm) was introduced that allows to derive a functional relationship between the measurements and shape parameters. We presented particular \rhms\ for elliptic and free-form star-convex shapes and derived measurement equations for which   standard Gaussian state estimators can be used. Nevertheless, estimating a detailed star-convex shape approximation is only possible when the measurement noise is rather low compared to target extent and enough measurements per time step are available. If this not the case, a basic shape such as an ellipse is more suitable.  Hence, a mechanism for  adapting the complexity of the used shape description is desired.  The   capability of estimating free-form shape approximations paves the way for new  applications, e.g., classification based on the shape and group splitting detection.

\bibliographystyle{IEEEtran}
\bibliography{Literatur,ISASPublikationen,ISASPublikationen_laufend}




\end{document}